\documentclass[fleqn,usenatbib]{mnras}

\usepackage{newtxtext,newtxmath}
\usepackage[T1]{fontenc}
\usepackage{ae,aecompl}

\usepackage{graphicx}	% Including figure files
\usepackage{amsmath}	% Advanced maths commands
\usepackage{amssymb}	% Extra maths symbols
\usepackage{booktabs}
\usepackage{eurosym}
\title[Extreme GeV Spectral Variability in IC 310]{\emph{Fermi}-LAT Observations of Extreme Spectral Variability in IC~310}

\author[Jamie A. Graham et al.]{
Jamie A. Graham,$^{1}$\thanks{E-mail: jamie.a.graham@durham.ac.uk}
Anthony M. Brown,$^{1}$
and Paula M. Chadwick$^{1}$
\\
$^{1}$Department of Physics and Centre for Advanced Instrumentation, University of Durham, South Road, Durham DH1 3LE, United Kingdom
}

\date{Accepted XXX. Received YYY; in original form ZZZ}

\pubyear{2015}

\begin{document}
\label{firstpage}
\pagerange{\pageref{firstpage}--\pageref{lastpage}}
\maketitle

\begin{abstract}
We investigate the physical mechanisms of high-energy gamma-ray emission from the TeV-emitting misaligned active galactic nucleus IC~310.
Eight years of data from the \emph{Fermi} Large Area Telescope (\emph{Fermi}-LAT) between 100~MeV and 500~GeV are reduced and analysed to study the temporal and spectral characteristics of IC~310. Point spread function-partitioned instrument response functions are used to improve the resolvability of IC~310 from nearby NGC 1275. Systematic effects due to this choice of instrument response functions and the proximity of NGC 1275 are investigated.
We find strong spectral variability and detect the hard flaring state of IC~310 along with a previously undiscovered soft state in quiescent periods, and the first detection with \emph{Fermi}-LAT below 1~GeV.
This represents a shift in peak Compton energy of more than five orders of magnitude. Possible interpretations are discussed, but we lack the instantaneous sensitivity with \emph{Fermi} to probe the underlying physics.
\end{abstract}


%%%%%%%%%%%%%%%%%%%%%%%%%%%%%%%%%%%%%%%%%%%%%%%%%%

%%%%%%%%%%%%%%%%% BODY OF PAPER %%%%%%%%%%%%%%%%%%

\section{Introduction}
The active galaxy IC~310 has shown substantial flux variability from GeV to TeV energies on time-scales around $20\%$ of its central black hole's radius \citep{aleksic_rapid_2014}. First detected at TeV energies during a MAGIC study of nearby NGC~1275 \citep{mariotti_magic_2010, aleksic_detection_2010}, IC~310 is a strong probe of the gamma-ray emission processes around galaxy cores.

Originally classified as a head-tail radio galaxy from its radio morphology \citep{sijbring_multifrequency_1998}, more recent observations with the European VLBI Network have shown blazar-like characteristics in the inner jet. \citet{schulz_evn_2015} used the absence of a detectable counter-jet to place an upper limit of $20^\circ$ on the viewing angle. Small angles are also ruled out to constrain the deprojected length of the jet. Its intermediate jet angle has led to IC~310's classification as a radio galaxy along with M87, Centaurus A and NGC~1275 in \emph{Fermi}-LAT source catalogues \citep{2FGL,acero_fermi_2015,gasparrini_3rd_2015}, all of which are also Very High Energy (VHE; $E_\gamma > 100$ GeV) emitters.

\citet{neronov_very_2010} confirmed the presence of IC~310's emission above 100~GeV using two years of \emph{Fermi} Large Area Telescope (\emph{Fermi}-LAT) data. During VHE outbursts in October 2009 and February 2010, the \emph{Fermi}-LAT detector in conjunction with uncontemporaneous X-ray observations constrained IC~310's broadband spectral energy distribution (SED) \citep{aleksic_rapid_2014}. These X-ray observations found substantial spectral variability, with the X-ray spectral index ranging from 1.76 to 2.55 throughout piecemeal observations over 4 years \citep{aleksic_rapid_2014}. Given the hard emission at TeV energies and above, IC~310 has now been re-classified as an AGN of unknown type \citet{Rieger_Levinson}, with the synchrotron component peaking in the X-ray regime (though not directly observed). The peak of the VHE emission was not found within the energy regime measured by MAGIC (when EBL deabsorbed) in either the high- or low-activity state by \citet{aleksic_rapid_2014}, a difference characterised by an order of magnitude in flux. With additional MAGIC observations outside of the flaring period, \citet{ahnen_first_2017} found a softer state as yet undetected with \emph{Fermi-LAT}.

%$[POINTLESS PARAGRAPH At optical and UV wavelengths, the host galaxy's thermal emission dwarfs IC~310's non-thermal emission by around two orders of magnitude. IC~310 is a high-peaked BL Lac object according to its high-energy emission and this dominance by thermal emission would be surprising but for our off-axis viewing angle. While IC~310 is not an archetypal HBL such as 1ES~1426+438 \citep{leonardo_multiwavelength_2009}, the identification is apt given current evidence, though further identity crises would not be concerning.]

Spectral variability within AGN has been observed at X-ray \citep{xrayspecvar} and gamma-ray wavelengths \citep[section 4.2]{BrightBlazarsFermi}. During AGN flaring periods a strong correlation between spectral hardness and flux is often observed, commonly referred to as `harder when brighter' behaviour \citep[e.g.][]{Anthony}. This spectral variability can be explained in simple homogeneous acceleration and cooling scenarios as shown by \citet{SpecVarCooling}.

IC~310's broadband spectral energy distribution (SED) has been modelled to some success using synchrotron self-Compton (SSC) models at varying assumed jet viewing angles. \citet{ahnen_first_2017} found that a single-zone SSC model could not explain the TeV flare due to Klein-Nishina suppression at such high energies. It also found that the second hump's peak had moved more than two orders of magnitude in energy during the TeV flaring period. Further modelling with multi-zone leptonic SSC models and hadronic components suggests that purely leptonic models may be disfavoured, as they require minimum Lorentz factors greater than $10^5$ \citep{Hadronic}.

In this paper we take advantage of the improvements in effective area and angular resolution afforded by \emph{Fermi}'s PASS 8 data to investigate IC~310's gamma-ray properties. In particular, we analyse the first eight years of \emph{Fermi}-LAT observations, utilising an improved point spread function (PSF) and updated analysis tools. Section~\ref{sec:NewSource} describes the data selected for our analysis. We found that several checks for systematic effects caused by our assumptions and reliance on the third \emph{Fermi} point source catalog (3FGL, \citet{acero_fermi_2015}) were required, which are performed in Sect.~\ref{sec:Results} and Sect.~\ref{sec:Systematics}. We discuss the ramifications of such a model in Sect.~\ref{sec:Discussion} and present our conclusions in Sect.~\ref{sec:Conclusion}.

\section{Data Reduction}\label{sec:NewSource}

\begin{table*}
\centering
\caption{Summary of the new sources with $\text{TS}>25$ found in our ROI. $^\text{a}$ PS J0355.3+3910 was no longer significant (TS<25) when an analysis centred on this source was performed. $^\text{b}$ While not in the 3FGL, this source was identified as 4C+50.11 in \citet{carpenter_fermi_2014}}
\begin{tabular}{lllll}
\hline
Point Source & RA (J2000) & Dec. (J2000) & Offset & TS \\ 
 & (degrees) & (degrees) & (degrees) & \\
\hline
PS~J0312.8+4121 & 48.21 & 41.35 & 0.716 & 42.07 \\
PS~J0355.3+3910 & 58.82 & 39.16 & 7.680 & 39.79$^\text{a}$ \\
PS~J0302.5+3353 & 45.63 & 33.88 & 7.958 & 51.36 \\
PS~J0257.1+3360 & 44.28 & 33.99 & 8.289 & 30.12 \\
PS~J0344.6+3433 & 56.15 & 34.55 & 8.725 & 36.32 \\
PS~J0233.0+3740 & 38.25 & 37.66 & 9.181 & 40.86 \\
PS~J0410.7+4218 & 62.68 & 42.30 & 10.107 & 74.66 \\
PS~J0344.2+3202 & 56.04 & 32.04 & 10.795 & 34.55 \\
PS~J0359.4+5052 & 59.86 & 50.87 & 12.055 & 359.99$^\text{b}$ \\
\hline

\end{tabular}
\label{tab:NewSources}
\end{table*}

Approximately eight years of \emph{Fermi}-LAT data from 2008-08-04 15:43:38 UTC until 2016-08-08 06:01:06 UTC (239557417 to 492328870 MET) are used in this analysis. All analysis was performed using version \texttt{v10r0p5} of the \emph{Fermi}-LAT Science Tools, with \texttt{P8R2} events and instrument response functions (IRFs). The $20^\circ$ region radially surrounding IC~310 (RA, Dec (J2000): 49.16458$^\circ$, 41.32947$^\circ$) was extracted from the \emph{Fermi}-LAT data server with energies in the range $100\,$MeV--$500\,$GeV. The size of this region ensures that low-energy photons from nearby sources were entirely contained within the region of interest (ROI). We performed a standard binned \emph{Fermi}-LAT analysis using the python tool \emph{fermipy}\footnote{\href{http://fermipy.readthedocs.io/en/latest/}{http://fermipy.readthedocs.io/en/latest/}}, keeping all photons with event class \texttt{SOURCE} or better (\texttt{evclass=128}). A $90^\circ$ zenith cut was applied along with the time selection filter \texttt{(DATA\_QUAL>0)\&\&(LAT\_CONFIG==1)}. The energy range is split into 10 logarithmically spaced bins per decade for analysis.

The \texttt{PSF} event types (\texttt{evtype=4,8,16,32}) were used to apply the appropriate IRF and isotropic background model to the data according to their PSF reconstruction quality. It was decided to work with the \texttt{PSF} IRFs and a summed likelihood method due to the proximity of IC~310 to NGC~1275, a bright gamma-ray source only $0.6^\circ$ away. These IRFs provide a far better ability to discriminate between photons originating from NGC~1275 and IC~310 but result in an increased systematic error on the effective area of the LAT ($20\%$ as opposed to $10\%$ when using the \texttt{FRONT} and \texttt{BACK} event types at the highest energies).

We use the standard definition of the likelihood ratio test statistic (TS) to assess the significance of a source:
\begin{equation}
TS = 2(\log(\mathcal{L}_1) - \log(\mathcal{L}_0)),
\end{equation}
where $\mathcal{L}_1$ and $\mathcal{L}_0$ are the maximum likelihood of the best-fit sky model with and without the source in question. This statistic is approximately distributed as a $\chi^2_1$ when only the normalisation of the source is left to vary using an assumed power-law spectrum \citep{mattox_likelihood_1996}. For all $\sigma$ values quoted in this paper, we assume that this approximate distribution is correct and thus use the simplification that a model is preferred to $n\sigma$ where $n = \sqrt{\text{TS}}$ unless specified otherwise. Particular care was taken to refit the normalisation of NGC~1275 when evaluating the TS of IC~310 as its proximity caused a substantial systematic error if this was neglected (as might be introduced when na{\"i}vely using the \texttt{Ts} function within the \emph{Fermi}-LAT python tools).

The 4-year \emph{Fermi}-LAT point source catalog (3FGL, \citet{acero_fermi_2015}) was used to seed a preliminary sky model along with the galactic diffuse model \texttt{gll\_iem\_v06.fits} and the appropriate isotropic diffuse model of the form \texttt{iso\_P8R2\_SOURCE\_V6\_PSF(x)\_v06.txt}, where x denotes the PSF classes 0 to 3, before examining the residual map for undetected point sources. The analysis itself was performed using the \emph{fermipy} routine \texttt{optimize} followed by the deletion of weak (TS~$<1$) sources from the sky model and refitting of bright (TS~$>25$) sources. The shape parameters of all bright sources are allowed to vary except that of IC~310, which is kept fixed to its 3FGL shape. In addition to the detailed analysis of IC~310 and NGC~1275 presented in this paper, nine new sources with $\text{TS}>25$ were discovered within the ROI using an iterative TS map peak finding technique. Their properties are listed in Table~\ref{tab:NewSources}.

Power law and log parabolas are used as phenomenological spectral models to perform the maximum likelihood fit as a function of energy in this work. The power law model is defined as:
\begin{equation}
\frac{dN}{dE} = N_0 \left( \frac{E}{E_0} \right)^{-\Gamma},
\label{eqn:pl}
\end{equation}
where $N_0$ is the overall normalisation factor to scale the observed brightness of a source, $E_0$ is a scale energy (held fixed at its 3FGL value), and $\Gamma$ is the power law index controlling the hardness of the source.
Similarly, the log parabola model:
\begin{equation}
\frac{dN}{dE} = N_0 \left( \frac{E}{E_0} \right)^{-\left(\alpha + \beta \log\left(E/E_0\right) \right)},
\label{eqn:lp}
\end{equation}
has an overall normalisation factor ($N_0$) and fixed scale energy ($E_0$), but the spectral scaling contains an energy dependent term $\beta$ along with an energy-independent term $\alpha$.

\section{Results}\label{sec:Results}

\subsection{IC~310}

\begin{figure*}
\includegraphics{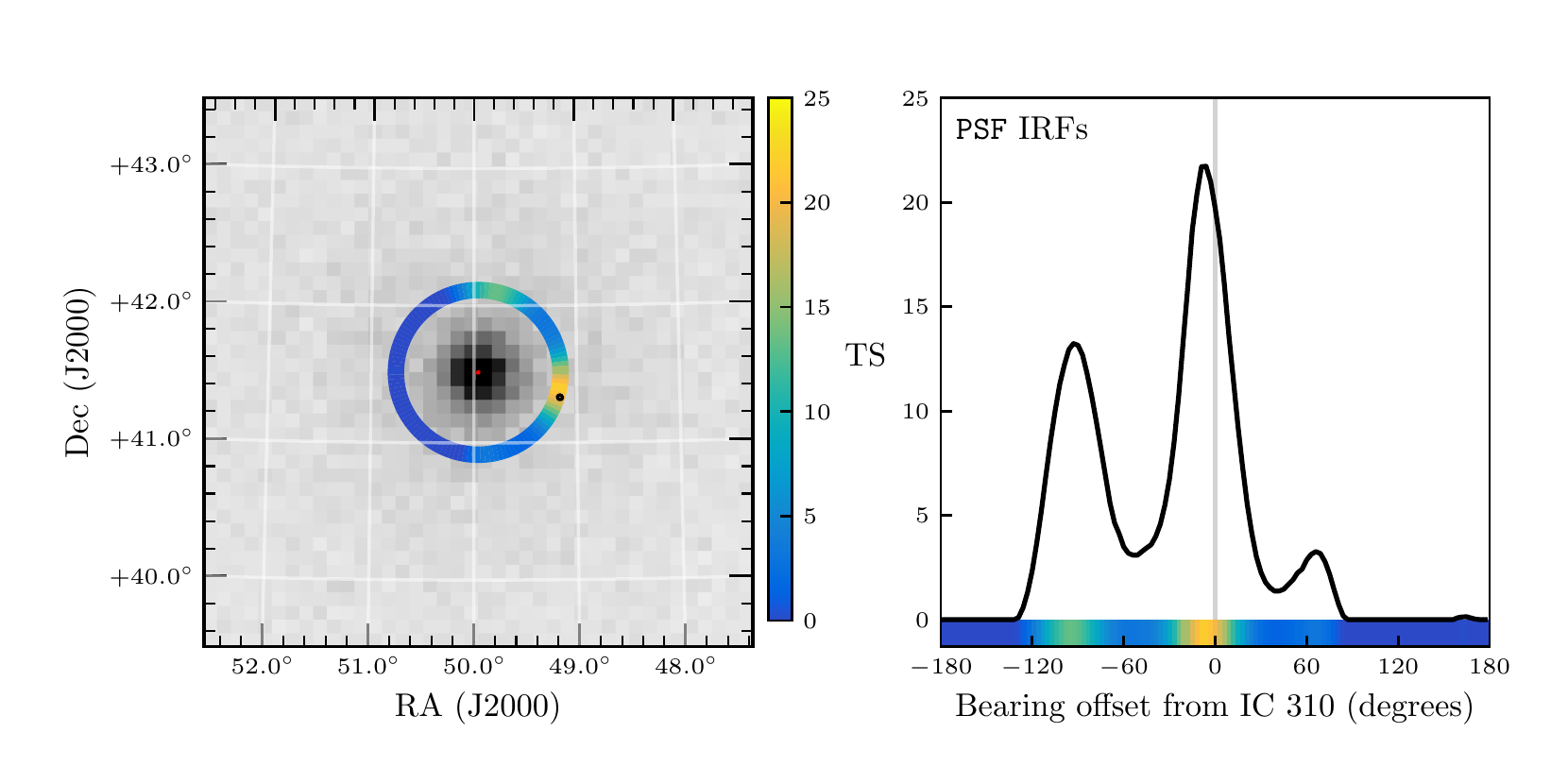}
\caption{Left: The \emph{Fermi}-LAT counts map centred upon NGC~1275 with a 1$^\circ$ radius using 0.1$^\circ$ bins. The annulus is coloured according to the TS of a point source with index $\Gamma=2.45$ at this position in an analysis between 578.5~MeV and 3~GeV using a composite \texttt{(PSF0 + PSF1 + PSF2 + PSF3)} likelihood. The red ellipse represents the localised position of NGC~1275 from Sect.~\ref{sec:Localisation} and the black ellipse represents the localised position of IC~310 obtained from the 3FGL. Right: The annulus TS values unwrapped and plotted as a function of bearing relative to IC~310. The corresponding colourmap value is plotted underneath.}
\label{fig:Circle}
\end{figure*}

Using the 3FGL spectral parameters of IC~310, we find the source is detected with a TS of 64.4 ($\sim8\sigma$ with 1 degree of freedom) between 100~MeV and 500~GeV. When allowing the index of IC~310 to vary, the spectral index softens significantly (2.34$\sigma$)to $\Gamma = 2.34$ from the 3FGL value of $\Gamma=1.90$ (before the background is investigated further). By comparing the $2\times\Delta \log(\mathcal{L})$ when allowing the index to vary, we find that the softer index is preferred with $\text{TS}=23.6$, which corresponds to a superior fit to the data with a preference of $\sim4.8\sigma$. This softer spectrum contrasts with the previous spectral studies of IC~310 \cite{neronov_very_2010}, though we use an expanded data set.

Given that we are observing significant low-energy emission from IC~310 with a bright source with strong spectral curvature only $0.6^\circ$ away, the model of IC~310 may be attributing some of NGC~1275's emission to itself at low energies. The PSF of the \emph{Fermi}-LAT instrument is on the order of $5^\circ$ at $100\,$MeV. To investigate this possibility, in Figure~\ref{fig:Circle} we test the TS at 120 positionings at various bearings from NGC~1275 on a circle with radius $0.6^\circ$ around the 3FGL position of NGC~1275 (corresponding to a change of $3^\circ$ in bearing between each test point). We constrain the energy range for this analysis to ensure all signal is associated with the low-energy behaviour. We choose a lower energy bound of 578.5 MeV by calculating the 95\% containment region on PSF3 class photons calculated for our ROI using the \texttt{gtpsf} tool, and choose 3 GeV as the upper energy bound as the approximate upturn of the SED found in Section~\ref{sec:NewSource}. We find a very strong peak (TS=21.73) in the direction of IC~310, along with two smaller peaks (TS = 13.2 and 3.3) at approximately (RA, Dec): (49.403$^\circ$, 41.948$^\circ$) and (50.279$^\circ$, 40.964$^\circ$) respectively. While this strong peak can be interpreted as the low-energy emission of a quiescent IC~310, the provenance of the two weak peaks is unclear. Given their angular offset (slightly less than $180^\circ$) and the fact that extended emission has been  observed in the radio galaxies Centaurus A and Fornax A \citep{CentaurusA, FornaxA}, we investigate the prospect of extended emission in Sect. 4.1. Dark matter present in the Perseus cluster could cause some extended emission, but the peaked nature of this emission contrasts with the large-scale smooth emission expected from DM annihilation in galaxy clusters \citep[see e.g.][Fig. 1]{Han_DM}.

\subsection{Integrated spectral energy distribution}\label{sec:OtherSource}

\begin{table*}

\centering
\caption{A summary of the sky models tested in Sect.~\ref{sec:NewSource}. $^\text{a}$ The relative number of degrees of freedom ($k$) in each model as compared to model A. $^\text{b}$ These values are approximate, and are discussed further in Sect.~\ref{sec:NewSource}. $^\text{c}$ As the $\log(\mathcal{L})$ values returned from the \emph{Fermi} tools is only relatively correct due to the suppression of model-independent terms, we show only the change in AIC relative to model A.}
\begin{tabular}{lllllll}
\hline
Name & Model & $-\log(\mathcal{L})$ & $\Delta k$$^\text{a}$ & TS$^\text{b}$ & Signif.$^\text{b}$ & AIC$^\text{c}$\\
& & & & & $(\sigma)$ &\\
\hline
A & IC~310 Catalog Index & 1886808.45 & 0 & - & - & -\\
B & IC~310 Power Law (PL) & 1886805.72 & 1 & 5.47 & 2.34 & -3.46\\
C & IC~310 Log Parabola (LP) & 1886798.19 & 2 & 20.52 & 4.14 & -16.52\\
D & IC~310 (PL) + PS (PL, Index=2.0) & 1886791.06 & 4 & 34.79 & 5.02 & -26.78\\
E & IC~310 (PL) + PS (PL, Index Free) & 1886789.63 & 5 & 37.66 & 5.05 & -27.64\\
F & IC~310 (LP) + PS (PL, Index=2.0) & 1886784.43 & 5 & 48.04 & 5.91 & -38.04\\
G & IC~310 (LP) + PS (PL, Index Free) & 1886784.20 & 6 & 48.50 & 5.74 & -36.5\\
\hline
\end{tabular}
\label{tab:Models}
\end{table*}

\begin{figure}
\includegraphics{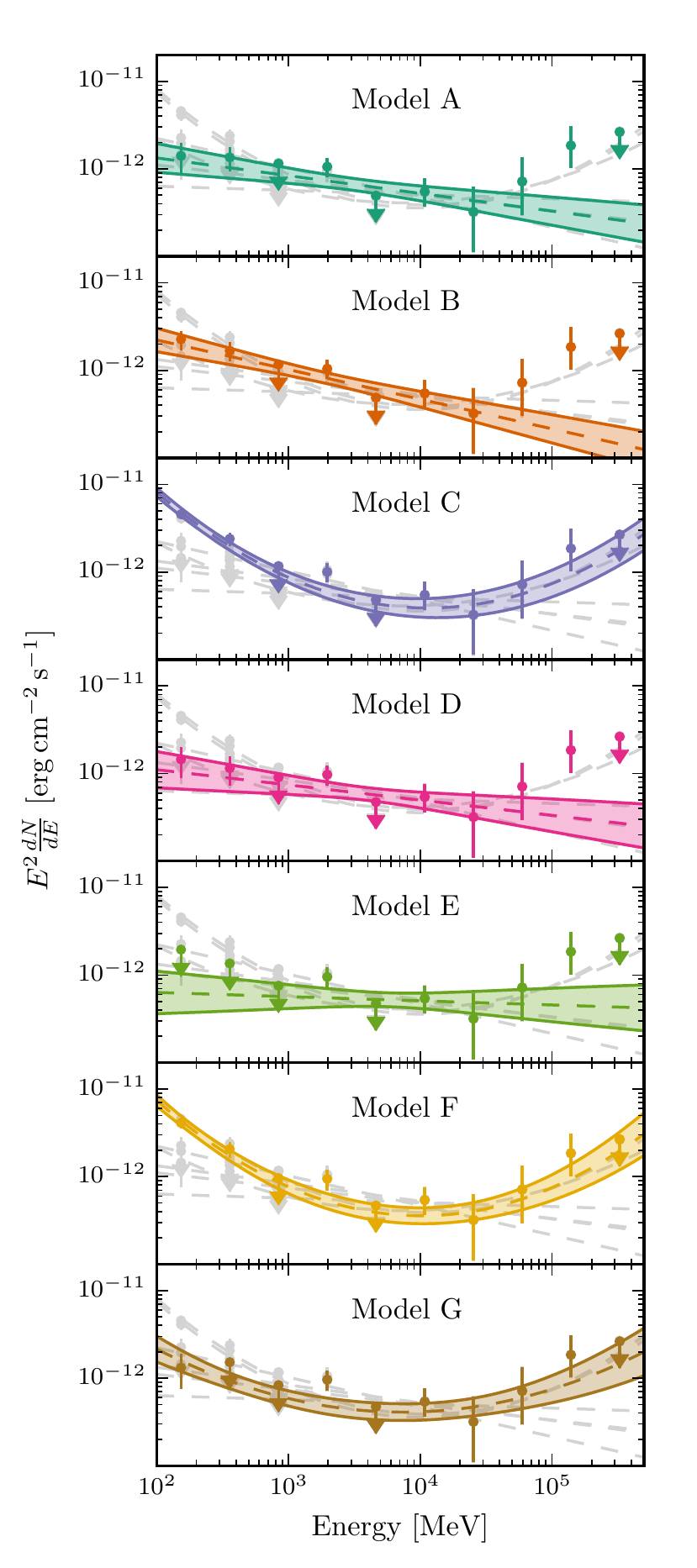}
\label{fig:SED}
\caption{The variation of the SED of IC~310 for each model considered in Table~\ref{tab:Models}. Each SED was calculated by profiling the normalisation of IC~310 in 10 logarithmically-spaced bins between 100$\,$MeV and 500$\,$GeV. It is readily seen that only the 3 low energy bins are significantly affected by this modelling, the source of our scepticism in Sects. \ref{sec:Extension}, \ref{sec:Localisation} and \ref{sec:OtherSource}. Upper limits are shown if the flux density in an energy bin is consistent with 0 within 2$\sigma$. }
\end{figure}

Due to the proximity of the source PS~J0312.8+4121, we optimise a series of different models that could fit the emission in the surrounding region of IC~310. By inspecting the SED of IC~310 one discovers an upturn at low energies. In fact, this so heavily weights the SED that, when freed, the best-fit index has a value of $-2.34$. Given that IC~310 is a TeV-emitting source this would be very unusual, though spectral hardenings towards the high end of the \emph{Fermi}-LAT energy range are not unheard of \citep[see][]{brown_discovery_2017}. MAGIC measured its spectrum consistently emitting with an index of -1.9 over several nights, and hardening to -1.3 during its brightest flare \citep{eisenacher_glawion_black_2015}. We thus consider a log parabolic model which finds a negatively curved spectrum with a $TS_{curve}$ of 20.52 when compared with the catalogue index (with varying normalisation), and $TS_{curve}=15.05$ when compared to the softer index. In each case, the scale energy of the log parabola was kept fixed at the lower end of the observed energy range as recommended by \cite{massaro_log-parabolic_2004}.

We provide a comparison of a set of feasible hypotheses in Table~\ref{tab:Models}. 

The models considered are as follows:
\begin{itemize}
\item Model A: The model for IC~310 is a power law with the index fixed to its 3FGL value. This serves as the reference model for our work compared to the null hypothesis, which in this case is the 3FGL.
\item Model B: The spectral index parameter for IC~310 is varied in addition to its normalisation.
\item Model C: Spectral curvature is introduced by substituting a log parabola spectral model for the power law description of IC~310. These models are nested and so can provide a direct comparison. The $\Delta \log(\mathcal{L})$ between models C and B is the $\text{TS}_{\text{curve}}$ as defined in the 2FGL \citep{2FGL}, though distinct from \texttt{Signif\_Curve} as we do not take into account systematic uncertainties.
\item Model D: We now add the point source PS~J0312.8+4121 to model B. The spectral model is assumed to be a power law with spectral index 2.0. In this model, we have also allowed the spectral index of IC~310 to vary. This results in 4 additional degrees of freedom from model A, namely the spectral index of IC~310, and the normalisation and position (RA and Dec.) of PS~J0312.8+4121. The 3 parameters associated with PS~J0312.8+4121 are not nested (there is a degeneracy when the normalisation of the source is zero), however, assuming the calibration performed by \citet{mattox_likelihood_1996} persists, all significance test values remain accurate.
\item Model E: The spectral index of PS~J0312.8+4121 is allowed to vary from model D. The additional free parameters are now the two spectral indices, and the normalisation and position of PS~J0312.8+4121.
\item Model F: IC~310 is now replaced with the log parabola spectral model from model C, along with a fixed power law spectral model for PS~J0312.8+4121 as in model D.
\item Model G: The spectral index of PS~J0312.8+4121 in model E is now allowed to vary. This was the first additional parameter that did not result in an overall increase in significance.
\end{itemize}
It should be noted that only models (A, B, C) and (D, E, F, G) are truly comparable in terms of $TS$. This is because the position was not optimised in each case for models D, E, F, and G. As the parameter was not truly free, the different TS values are only approximations. As the models may not be fully optimised, the TS may be lower than the fully optimised values. We can at least be assured that the optimal value would result in larger TS values, not smaller ones (when directly compared to model D). Comparisons between models E, F, and G assume the position of PS~J0312.8+4121, and may be larger or smaller when the position is minimised (unlike comparisons with model D), as the relative changes in $\log(\mathcal{L})$ are not calculated.

The summary in Table~\ref{tab:Models} shows model F to be the most likely model given the observed data. In each case the sky model was optimised and the log likelihood extracted. Each additional degree of freedom must decrease the $\log(\mathcal{L})$ by 1.0 to improve the significance, the relationship defining the Akaike Information Criterion \citep[AIC,][]{akaike_new_1974}. We also include the AIC as an independent test of significance assuming that approximate nesting is insufficient. The significance (in units of $\sigma$) is calculated as:
\begin{equation}
\text{Significance} = \sqrt{2} \hspace{0.1cm} \text{erf}^{-1}(F_{\chi^2}(TS, k)),
\end{equation}
where erf$^{-1}$ is the inverse error function, $F_{\chi^2}(x,k)$ is the cumulative distribution function of a $\chi^2$-distributed statistic at a value of $x$ with $k$ degrees of freedom, and $\sigma$ refers to one standard deviation of a normal distribution.

The AIC is defined as:
\begin{equation}
AIC \triangleq 2k - 2\log(\mathcal{L}),
\end{equation}
where $k$ is the number of degrees of freedom in the model, and $\mathcal{L}$ is defined as above. The model with the minimum AIC value is the preferred model. Appendix~\ref{App:Parameters} contains the optimised spectral parameter values for each model tested in this Section.

\begin{figure}
\includegraphics{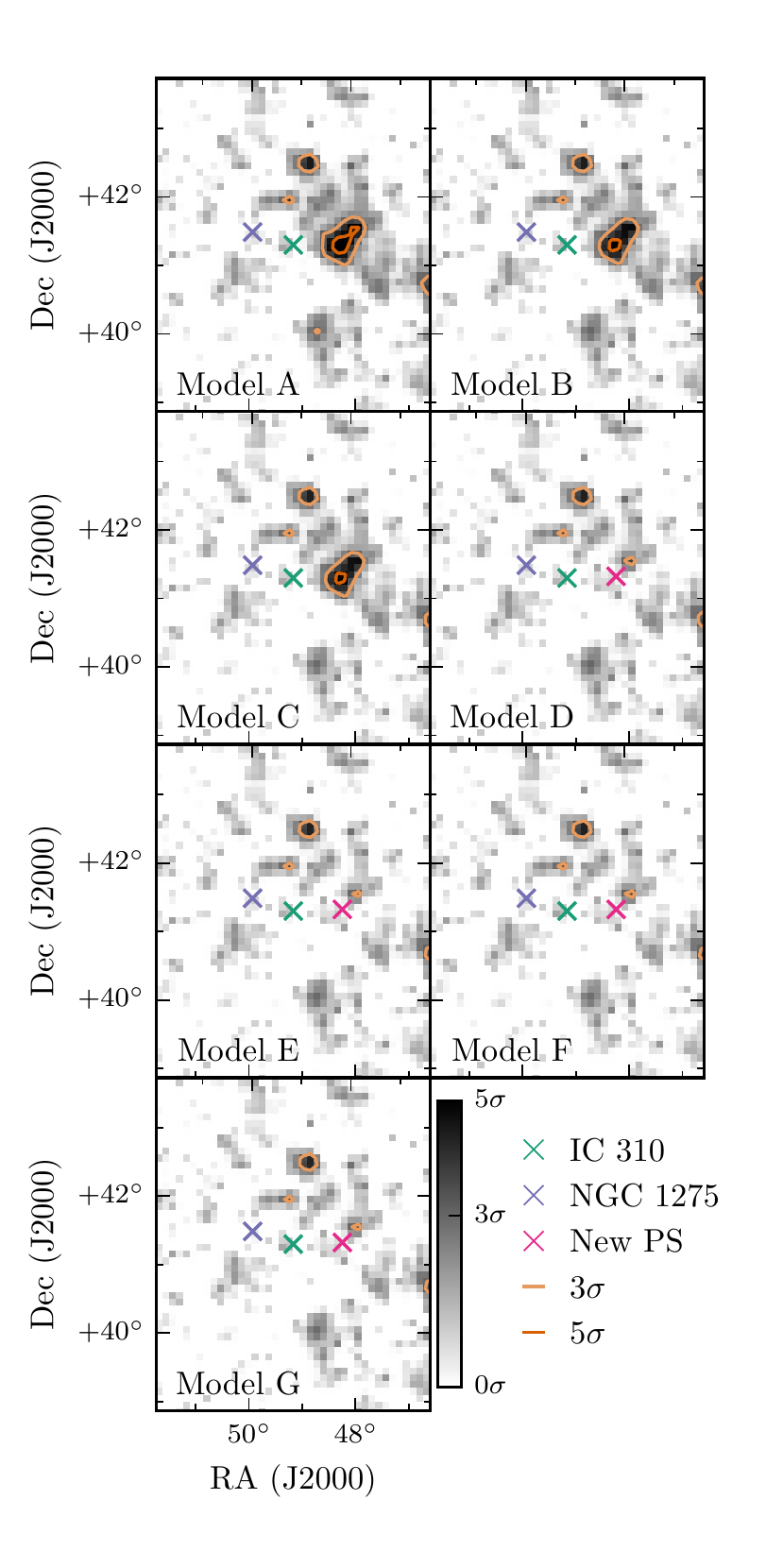}
\caption{A TS map of the ROI in a 3$^\circ$ radius around IC~310, showing the relevant point sources for each model assessed in Sect.~\ref{sec:OtherSource}. The statistical significance of PS~J0312.8+4121 can be seen as a peak before a point source model is included from model D onwards. Other potential point sources manifesting as smaller peaks can be seen, but we limit ourselves to a $5\sigma$ threshold for new point sources.}
\label{fig:TSMaps}
\end{figure}

The best-fit model (both when assuming nesting and the AIC) is shown to be that with IC~310 with negative $\beta$, preferred by $\sim$3$\sigma$ over a power law with index 2.01. In all cases, a negatively-curved spectrum is preferred for IC~310 over any models without such curvature. Curvature in IC~310 is preferred to a softer spectrum of PS~J0312.8+4121 by calculating the difference between models D and E, and models F and E. It should be noted that with its index left to vary, PS~J0312.8+4121 does not achieve 5$\sigma$ significance (a TS of 28.744) with 2 degrees of freedom between models G and C. Due to these facts, and the overall significance of each model, we find that model F allows the greatest overall significance given the number of parameters introduced. Freeing the index of PS~J0312.8+4121 between models F and G is shown to be overfitting, reducing the overall significance of the model in both statistical metrics. The SED of IC~310 given each model is shown in Figure~\ref{fig:SED}. Similarly, we plot the TS map of the ROI for each of the models discussed in this Section in Figure~\ref{fig:TSMaps}, which shows a convincing case for localised excess emission in the region of PS~J0312.8+4121, and thus the necessity of including this source in our models.

\subsection{Temporal Evolution}\label{sec:Temporal}

\begin{figure}

\includegraphics{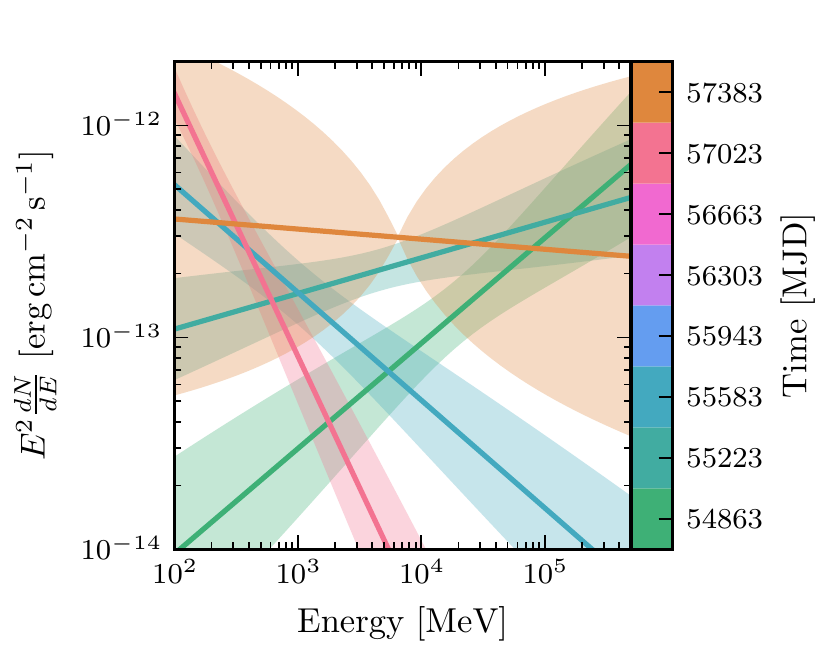}
  \caption{The SED for each annual bin, where each spectrum shown has a TS of at least 19.33 representing an approximate 4$\sigma$ cut defined with two degrees of freedom: normalization and spectral index. The shaded areas around each optimised model represent the $1\sigma$ flux uncertainty propagated from the uncertainties in spectral model parameters after the likelihood fit.}
  \label{fig:4_sigma}
\end{figure}

\begin{figure}
\includegraphics{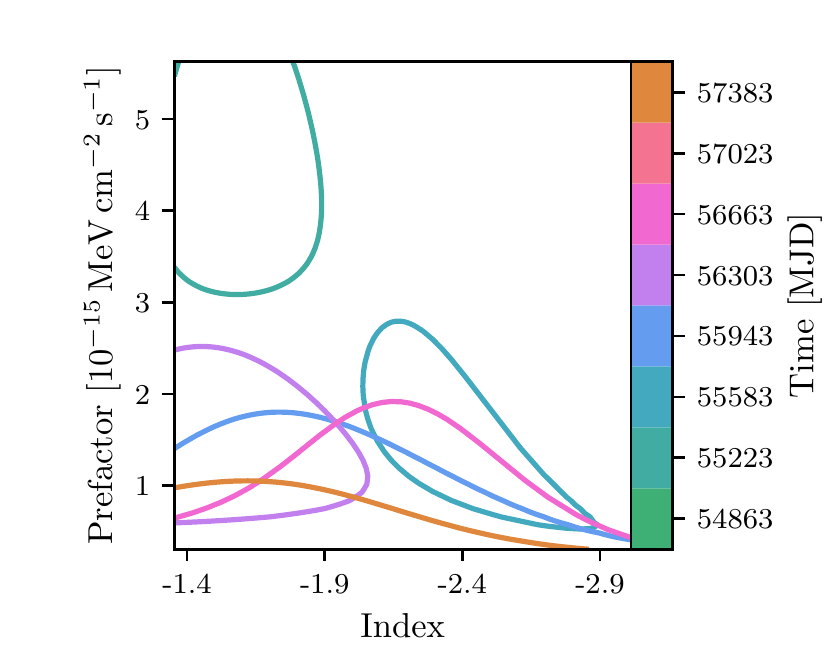}
  \caption{The $1\sigma$ uncertainty contour in the parameter space of a power law for each time bin assessed in Sect.~\ref{sec:Temporal}. The normalisation (but not spectral shape) of NGC 1275 has been left free to vary for each parameter combination. Bins whose optimised parameters are outside of this parameter space are excluded.}
  \label{fig:contour}
  \end{figure}

IC~310 is highly variable, not just in terms of flux, but also spectrally. This was noted by \citet{ahnen_first_2017} at VHE energies, but this should also be the case within the \emph{Fermi}-LAT energy range. To establish the temporal effects of integrating over quiescent and flaring periods observed by MAGIC, we can break the SED from Sect.~\ref{sec:NewSource} down into smaller time periods. Using 360-day bins, we re-fit Model E from Sect.~\ref{sec:OtherSource} to the data. This model was used in preference to Model F, as in only two of the eight time bins was Model F preferred to Model E at the 3$\sigma$ level. We plot these SEDs in Figure~\ref{fig:4_sigma}, which clearly shows the variation in power-law index with time. Once again, close attention was paid to ensure that NGC~1275 and PS~J0312.8+4121 are jointly optimised with IC~310. The parameter space is quite broad and some bins show only upper limits in flux. While the parameter space investigated is not large enough to fully encompass some of the time bins, it does show that in several periods IC~310 is incompatible with the harder emission previously detected.

\subsection{Investigating the spectral parameter space}

If we want to compare the spectra in each bin, it is more useful to look at a grid scan of the two spectral parameters in each case. This is performed by exhaustively calculating the likelihood given all combinations of parameter values. The parameter space plotted in Figure~\ref{fig:contour}, which shows a strong separation between hard and soft states. The limits on the parameter space grid scan were prescribed as half the minimum value of the parameter in the integrated fit and double the maximum value of the parameter in the integrated fit. In each case, the parameter value is set within the \texttt{pyLikelihood} instance of our total model fit and the ROI accordingly optimised with background sources free. This variation in spectral index is a convincing origin of the curved shape of the overall spectrum found in Sect.~\ref{sec:OtherSource}.

\begin{figure}
\includegraphics{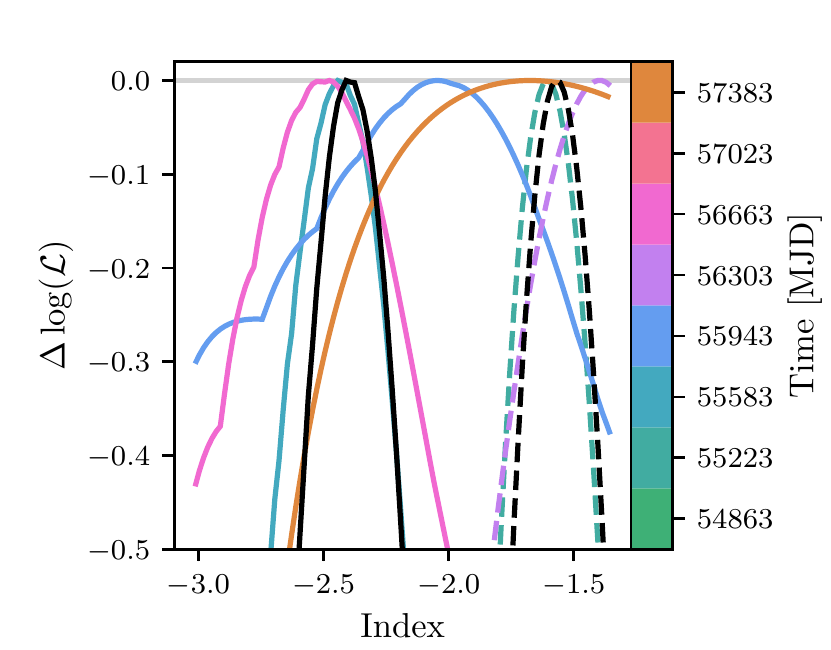}
  \caption{The likelihood space from Figure~\ref{fig:contour} profiling over normalisation. Solid profiles represent bins in which IC~310 was not detected by VHE telescopes (as determined from Table~\ref{tab:VHEDetections} and dashed lines represent VHE detected bins. The black lines represent the sum of the VHE detected and non-VHE detected profiles with the minimum subtracted, and thus the combined maximum. Artefacts are introduced due to the limited region over which the grid scan was performed and also due to the resolution chosen for the grid scan. Bins are chosen for comparison with Figure~\ref{fig:contour}.}
  \label{fig:Index}
  \end{figure}

Finding the maximum likelihood in each normalisation slice of this parameter space leads to a profile of the index parameter for IC~310. These profiles are shown in Figure~\ref{fig:Index}, and by summing the profiles with and without VHE flare detection from MAGIC (as shown in black), we find a distinct bimodality in index.

\begin{table*}
\centering
  \caption{Detailed information on the binnings assessed in Sect.~\ref{sec:Temporal}.}
  \begin{tabular}{lllll}\toprule
    Start [ISO (MJD)] & End [ISO (MJD)] & Index & TS$_\text{IC 310}$ & VHE Detection \\\midrule
    2001-08-04 (54682.66) & 2009-07-30 (55042.66) & -1.50 & 27.45 & -- \\
    2009-07-30 (55042.66) & 2010-07-25 (55402.66) & -1.84 & 57.78 & \citet{aleksic_detection_2010} \\
    2010-07-25 (55402.66) & 2011-07-20 (55762.66) & -2.71 & 25.82 & -- \\
    2011-07-20 (55762.66) & 2012-07-14 (56122.66) & -2.53 & 25.03 & -- \\
    2012-07-14 (56122.66) & 2013-07-09 (56482.66) & -1.65 & 9.68 & \citet{cortina_magic_2012}\\
             &          &       &       & \citet{ahnen_first_2017}\\
    2013-07-09 (56482.66) & 2014-07-04 (56842.66) & -2.55 & 19.37 & --\\
    2014-07-04 (56842.66) & 2015-06-29 (57202.66) & -2.05 & 20.37 & --\\
    2015-06-29 (57202.66) & 2016-06-23 (57562.66) & -1.57 & 26.84 & --\\\bottomrule
  \end{tabular}
  \label{tab:VHEDetections}
\end{table*}

\section{Systematic Checks}\label{sec:Systematics}

As in any analysis, there is a wide array of systematic effects that are introduced by our modelling methods. These range from specifics about how certain sources are treated in the sky model (both spectrally and spatially) to the choices we have made in Sect.~\ref{sec:NewSource} regarding the IRFs, cuts and large-scale sky models. In this section, we shall focus on a few of the areas which we believe to contribute the largest systematic uncertainties to our results in Sect.~\ref{sec:Results}. A further check for correlation between the flux in IC~310 and NGC~1275 due to mismodelling of NGC~1275 (and resultant spurious flux associated with IC~310) was investigated in Appendix~\ref{App:Degeneracy}, with no significant effect found.

\subsection{Extended emission from NGC~1275}\label{sec:Extension}

Source confusion between IC~310 and NGC~1275 presents a problem in our analysis given their $0.6^\circ$ separation. Especially at low energies where the LAT's PSF is larger, this may present problems if NGC~1275 is spatially resolvable (as mentioned in Sect.~\ref{sec:Results}, other radio galaxies have shown significant extension).

We thus search for extension in the point source of NGC~1275 by replacing it with a flat disc of various radii in the range $0.1^\circ$ -- $2.0^\circ$. The disc model is defined as:
\begin{equation}
Disc(x,y) =
\begin{cases}
\frac{1}{\pi\sigma^2} & \quad x^2 + y^2 \leq \sigma^2 \\
0   & \quad x^2 + y^2 > \sigma^2. \\
\end{cases}
\end{equation}
As shown by \citet{lande_search_2012}, both disc and Gaussian extension templates provide minimal bias when the size of the model is comparable to the size of the \emph{Fermi}-LAT point spread function. Again, following \citet{lande_search_2012}, we assess the significance of extension using the $\text{TS}_{\text{ext}}$ statistic:
\begin{equation}
\text{TS}_{\text{ext}} = 2\log(\mathcal{L}_{\text{ext}}/\mathcal{L}_{\text{ps}}),
\end{equation}
which was found to follow a $\chi^2_1/2$ distribution. It should be noted that this distribution was only validated up to a TS$_\text{ext}$ of 16 (a limitation due to statistics), thus we extrapolate no significance beyond $4\sigma$ for greater TS$_\text{ext}$ values. Radial extension from 0.01$^\circ$ up to $2.0^\circ$ were fitted, with a TS$_\text{ext}$ of 0.006 and an upper limit of 0.03$^\circ$ extension. This is well within the \emph{Fermi}-LAT point spread function, even at the highest energies, and rules out any significant extension in NGC~1275. There are no significant point sources (at the level of 5 standard deviations) immediately surrounding NGC~1275 that could bias this result.

At other wavelengths, far higher resolution studies of NGC~1275 could reveal the directionality of any existing extension. Previous radio observations (e.g. \citet{lister_mojave._2013}) finds that the inner jet axis is aligned in the North-South plane, incompatible with asymmetric extension in a westerly direction that could explain residual emission towards IC~310. This independent evidence suggests that extension is unlikely to be the cause of our signal.

\subsection{Localisation of NGC~1275}\label{sec:Localisation}

Alternatively, asymmetry as shown in Figure~\ref{fig:Circle} could be symptomatic of poor localisation of NGC~1275. We perform a maximum likelihood position estimation on NGC~1275 after removing IC~310 from the model. While this may bias us towards IC~310's position, the resulting error is far smaller than assuming IC~310's low-energy presence given the dominance of NGC~1275's photon statistics. Such a procedure finds a shift towards IC~310 of only 0.004$^\circ$ which is well within the 0.1$^\circ$ pixel size used in this analysis. 

Using this updated position, the preference of an integrated log parabola model over a power-law remains unaltered, though the localisation was not repeated in each time bin as \emph{a priori} we assume this will remain unaltered.

%Now that we can are safely convinced of this emission's pedigree, we can look to understand the processes that cause such a spectrum. As has been discussed, IC~310 is a violently variable AGN. Given the spectral curvature, it seems unlikely that this is contemporaneous emission. To determine whether this overall spectrum is, in fact, the convolution of spectral variability throughout the eight years of \emph{Fermi}-LAT data, we first calculate the lightcurve of IC~310 in yearly bins The data set is split into 365-day bins and the normalization of IC~310 and NGC~1275 are optimised in each bin, using the best-fit spectral characteristics found from the overall data set. In each bin, we fit only the normalisations of IC~310 and NGC~1275 (an incredibly variable object).

\subsection{Choice of IRFs}

As noted in Sect.~\ref{sec:NewSource}, in order to decouple IC~310's emission from that of NGC~1275, we chose to use the \texttt{PSF} IRFs. This choice increased our systematic error on the effective area, especially at the extreme ends of the \emph{Fermi}-LAT energy range, by around a factor of two. Our analysis is inherently spectral, so there is also an argument that the \texttt{EDISP} IRFs (in which the energy dispersion is minimised) would be appropriate. Alternatively, the combined \texttt{FRONT} and \texttt{BACK} IRFs could have provided a suitable middle-ground between energy dispersion and point spread function. In each case we will reanalyse this data while choosing a different set of IRFs and using the TS of IC~310's soft spectral state as our figure of merit (as the hard spectral state has been independently studied by \citet{neronov_very_2010}). Table~\ref{tab:IRFs} shows the effect on the significance of the soft state with each set of IRFs.

\begin{table}
\centering
  \begin{tabular}{cccc}\toprule
    IRF & Event Types & TS$_\text{Soft 1}$ & TS$_\text{Soft 2}$ \\\midrule
    PSF & (4 + 8 + 16 + 32) & 24.22 & 21.68 \\
    EDISP & (64 + 128 + 256 + 512) & 2.15 & $\sim 0$ \\
    FB & 3 & $2.58$ & $\sim0$ \\ \bottomrule
  \end{tabular}
  \caption{The significance of this result when varying the IRFs used in the analysis. The event types as defined by the \href{https://fermi.gsfc.nasa.gov/ssc/data/analysis/documentation/Cicerone/Cicerone_LAT_IRFs/IRF_overview.html}{\emph{Fermi}-LAT collaboration} are used in a composite likelihood analysis for PSF and EDISP classes, whereas the combined event class is used for the FB analysis. TS$_\text{Soft 1}$ refers to the third time bin (MJD: (55402.66, 55762.66)) and TS$_\text{Soft 2}$ refers to the seventh time bin (MJD: (56842.66, 57202.66)).}
  \label{tab:IRFs}
\end{table}

\begin{figure*}
\centering
\includegraphics{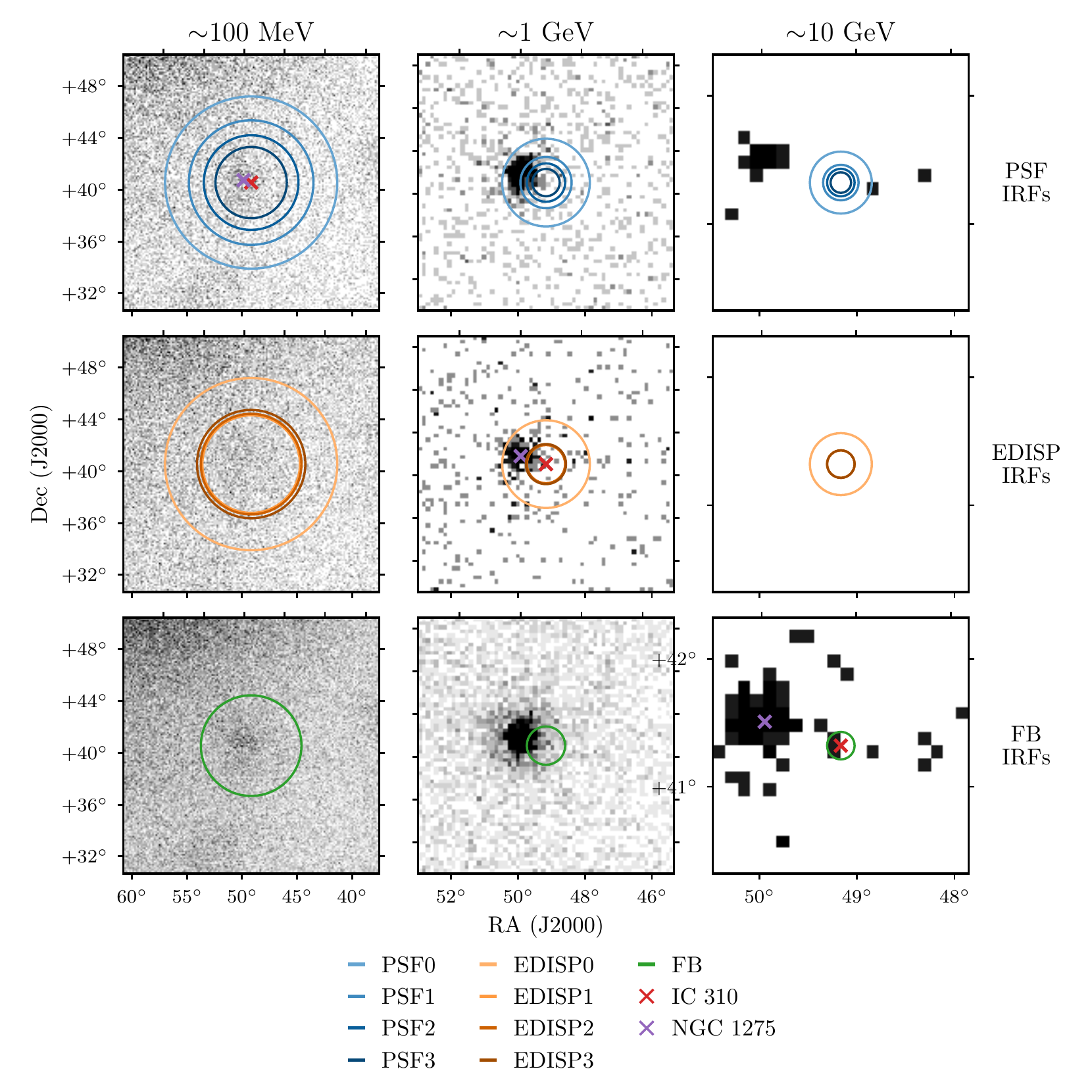}
\caption{The variation in PSF with IRF choice and energy. The data in the top row correspond to data partitioned into the PSF3 IRF class, the data in the middle row into EDISP3, and the data in the bottom row in the summed FB class (all photons would fall into this category). Data in the first column are binned in energy from 100 -- 126 MeV, data in the second column from 1 GeV to 1.26 GeV, and data in the final column from 10 -- 12.6 GeV. The image is zoomed between energy ranges for easier viewing, decreasing from $10^\circ$, to $3^\circ$, to $1^\circ$ with increasing energy. These data represent the first annual bin of our lightcurve. The circles plotted show the 95\% containment angles of the PSF of the \emph{Fermi}-LAT instrument using the IRFs as listed in the legend. Containment angles were calculated by numerically integrating the PSFs returned by the \texttt{gtpsf} tool within the \emph{Fermi} science tools. Asymmetry due to the fisheye effect and inclination angles to the LAT boresight is not considered. The positions of NGC~1275 and IC~310 are marked along the diagonal to show the scales relevant to our analysis.}
\label{fig:PSF}
\end{figure*}

\begin{figure}
\includegraphics{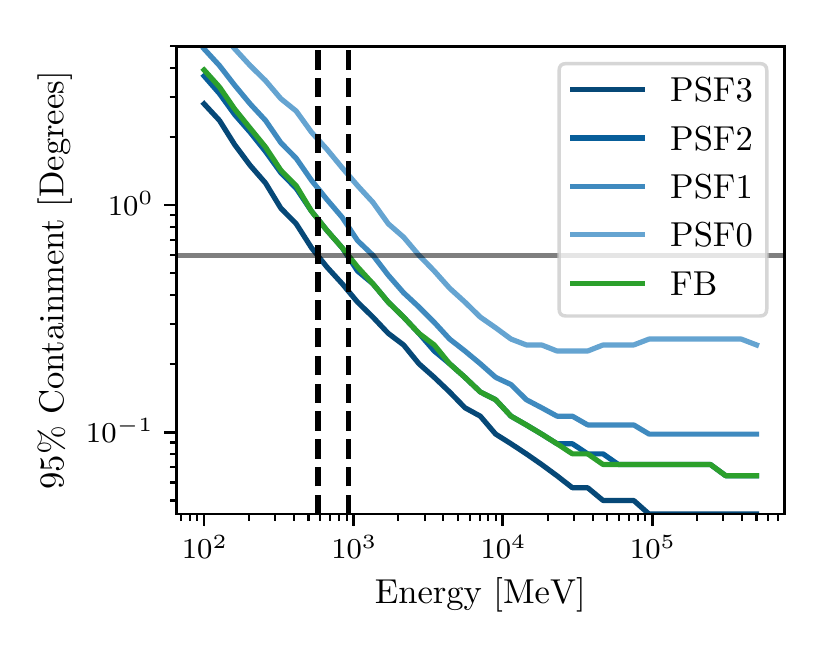}
\caption{95\% containment radii for our region of interest as a function of energy and instrument response function partition. Each line represents an individual IRF class, coloured as described in the legend. Vertical lines represent the 95\% containment regions of the PSF3 and FB classes, and the horizontal line represents 0.6$^\circ$, the separation of IC~310 and NGC~1275. Artefacts are due to the finite resolution used in the \texttt{gtpsf} tool used to calculate these values.}
\label{fig:PSF_2}
\end{figure}

The significance of the result drops significantly as the PSF of the set of IRFs increases. This is due to the inability to differentiate between IC~310 and NGC~1275 with a degraded PSF. While this vindicates our decision to use the PSF-partitioned IRFs, it does show that this analysis is sensitive to our choice of IRFs. This is not a cause for concern as the PSF partitioned IRFs have been well-validated, but should be noted in any case. As a visual representation of the PSF of the LAT instrument at various energies, Figure~\ref{fig:PSF} shows the variation of PSF (and data) with IRF choice. Figure~\ref{fig:PSF_2} further shows that the energy at which the 95\% containment radius exceeds 0.6$^\circ$ (the separation between NGC~1275 and IC~310) is severely reduced using the PSF3 IRFs. With the FB IRFs this energy is 928.6 MeV, but with PSF3 IRFs, this is reduced to only 578.5 MeV, almost a factor of two improvement.

\section{Discussion}\label{sec:Discussion}

We have reported the detection of a soft emission state in addition to the very energetic emission commonly associated with IC~310. While noted for these bright flaring periods, by inspecting the SED of IC~310 over the lifetime of \emph{Fermi} to date we find this soft emission is concentrated between periods in which strong flares have been observed.

\begin{figure*}
\centering
\includegraphics{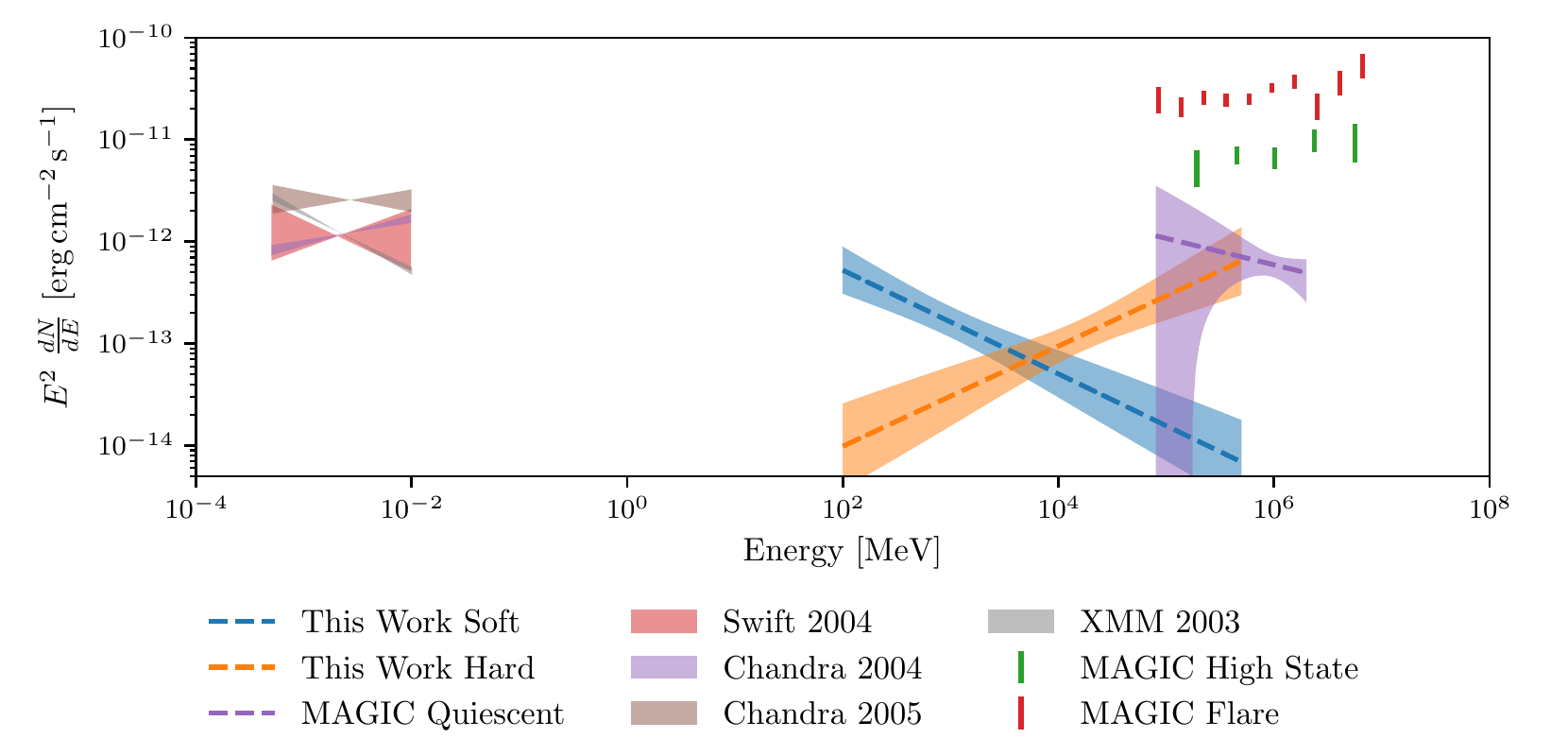}
\caption{The analysis from this work plotted with the data presented in \citet{aleksic_rapid_2014}. The `hard' spectrum plotted from this work represents the first bin of our analysis (MJD: 54682.66, 55042.66), and the `soft' spectrum represents the third bin (MJD: 55402.66, 55762.66). These data are not simultaneous.}
\label{fig:Broadband}
\end{figure*}

The two states, flaring and quiescent, are most recognisably distinct in their spectra. The peak of the inverse-Compton peak during flaring periods was undetected even above 10~TeV, while our analysis has found the inverse-Compton peak to be lower than 100~MeV during quiescent flux states under the assumption of a synchrotron self-Compton model. This represents an increase of more than 5 orders of magnitude in inverse-Compton peak energy on time-scales of a year. Similar spectral variance has been seen over multiple X-ray observations as shown in Figure~\ref{fig:Broadband}, which could represent the same behaviour for the synchrotron peak. 

Without simultaneous multiwavelength observations to constrain the broadband SED it is almost impossible to make conclusions as to how this bimodal behaviour can be modelled. Recent simultaneous observations by \citet{ahnen_first_2017} find that a simple SSC model can sufficiently explain the broadband behaviour of IC~310 using different viewing angle estimates. Interestingly, the spectral indices of the analysis presented in this paper and the index of the VHE quiescent emission detected by \citet{ahnen_first_2017} are consistent. However they predict a peak IC energy around 80~GeV, which would disagree with the downward-going spectrum detected here with \emph{Fermi}-LAT. It should be noted that the measurements presented in this work and in \citet{ahnen_first_2017} are not contemporaneous, so cooling of the electron population could explain this change in peak energy.

To gain simultaneous observations of the Compton and inverse-Compton peaks, X-ray instruments must integrate exposure on IC~310 over the year-long time periods necessary for significant detection using \emph{Fermi}. Assuming the SSC paradigm, we can make some physical inferences. The cooling time-scales of the electrons emitting gamma-rays at GeV energies are far shorter than the time periods (year-long bins) that we consider, and the flares could be explained (at least in the \emph{Fermi} energy range) by a Doppler factor increase due to changes in the jet. Without simultaneous broadband observations, however, these inferences remain speculation but can be explained by a plausible mechanism.

Whilst we have shown that the integrated spectrum of IC~310 does indeed show a negative $\beta$ parameter, this should not be interpreted as an intrinsic property of the electron energy distribution of IC~310. This would have to result from an upturn in energy spectrum of accelerated particles moving down the jet, an unlikely prospect at the level required for these results. This is unambiguously an artefact of time integration.

%Given the extremely short variability timescales associated with this source \citep{aleksic_black_2014} made possible by instruments with superior instantaneous sensitivity than the \emph{Fermi}-LAT instrument used in this analysis, it is safe to assume that integrating over 8 years of this data will introduce artefacts due to unresolvable temporal changes in the source's emission.

\section{Conclusions}\label{sec:Conclusion}

We have discovered a quiescent state in the \emph{Fermi}-LAT spectrum of IC~310, which is spectrally distinct from the flaring state that has been previously studied. Whilst an analysis of the 8-year data set results in a strongly curved spectrum (an upturn with increasing energy), we conclude that this is an artefact of integrating a hard, flaring state and a soft, quiescent state. Between the two spectral states, the power law index hardens from $\Gamma=2.71$ to $\Gamma=1.50$. 

\citet{ahnen_first_2017} found that the peak energy of the second SED hump extended from below 100~GeV to over 10~TeV, a jump of two orders of magnitude. This work pushes the peak energy of this hump below even 100~MeV, resulting in a jump of over 5 orders of magnitude in peak energy. 

To fully understand the emission processes responsible for IC~310's SED and these large changes in injected particle energy, it will be necessary to undertake further simultaneous broadband observations, ideally with good instantaneous sensitivity. Forthcoming instruments such as the Cherenkov Telescope Array \citep{CTA} observing simultaneously with \emph{NuSTAR} \citep{nustar} at hard X-ray wavelengths would substantially constrain the quiescent spectrum in preparation for follow-up of future flaring activity.

\section{Acknowledgements}
JG is supported by an STFC studentship, grant reference ST/N50404X/1. AMB and PMC are supported by the STFC Consolidated Grant, reference ST/P000541/1. This work has made use of the NASA/IPAC Extragalactic Database (NED), which is operated by the Jet Propulsion Laboratory, Caltech, under contract with the National Aeronautics and Space Administration. This paper makes use of publically available \emph{Fermi}-LAT data provided online by the \href{http://fermi.gsfc.nasa.gov/ssc/}{Fermi Science Support Center}.

%%%%%%%%%%%%%%%%%%%%%%%%%%%%%%%%%%%%%%%%%%%%%%%%%%



%\bibliographystyle{mnras}


%%%%%%%%%%%%%%%%%%%%%%%%%%%%%%%%%%%%%%%%%%%%%%%%%%

%%%%%%%%%%%%%%%%% APPENDICES %%%%%%%%%%%%%%%%%%%%%

\begin{appendix}\label{App:Parameters}

\section{Spectral parameters of IC~310}

The spectral parameters of each model evaluated in Sect.~\ref{sec:NewSource} are listed in Table~\ref{tab:Models_2}. In each case, the entire region of interest was independently fit after any change to the spectral or spatial model. Power law and log parabola models refer to the spectral models defined by Equations~\ref{eqn:pl} and~\ref{eqn:lp}.

\begin{table*}
\centering
\caption{The spectral characteristics of IC~310 in each of the models considered in Sect.~\ref{sec:OtherSource}.}
\begin{tabular}{l l l l l}
\hline \hline
\multicolumn{5}{c}{Power Law Models} \\
\hline
Model & Prefactor &Index      &Scale      &  \\
& (MeV$\,$cm$^{-2}\,$s$^{-1}$) & & (MeV) & \\ \hline
A & $(2.96\pm0.60)\times 10^{-13}   $ &$-1.9   $ &$6.42\times 10^{3}   $   &     \\

B & $(6.38\pm1.47)\times 10^{-13}$ &$-(2.34   \pm 0.163)  $ &$6.42\times 10^{3}   $   &     \\

D & $(8.13\pm1.57)\times 10^{-15}$ &$-(2.17   \pm 0.19)   $ &$6.42\times 10^{3}$   &     \\

E & $(7.87\pm1.59)\times 10^{-15}$ &$-(2.05   \pm 0.193)  $ &$6.42\times 10^{3}$   &     \\

\multicolumn{5}{c}{} \\\hline \hline
\multicolumn{5}{c}{Log Parabola Models} \\ \hline
Model & Prefactor  &Alpha       &Beta        &Eb       \\
& (MeV$\,$cm$^{-2}\,$s$^{-1}$) & & & (MeV) \\ \hline
C & $(5.4 \pm1.01)\times 10^{-13}$ & $2.66    \pm0.0793  $ & $-0.136  \pm0.0268  $ & $1\times 10^{3}   $ \\

F & $(4.78\pm1.41)\times 10^{-13}$ & $2.66    \pm0.11    $ & $-0.142  \pm0.0328  $ & $1\times 10^{3}   $ \\

G & $(3.66\pm0.812)\times 10^{-14}$ & $2.36    \pm0.0937  $ & $-0.0894 \pm0.0276  $ & $1\times 10^{3}   $ \\ \bottomrule

\end{tabular}
\label{tab:Models_2}
\end{table*}

\section{Degeneracy with NGC 1275}\label{App:Degeneracy}

\begin{figure*}
\centering
\includegraphics{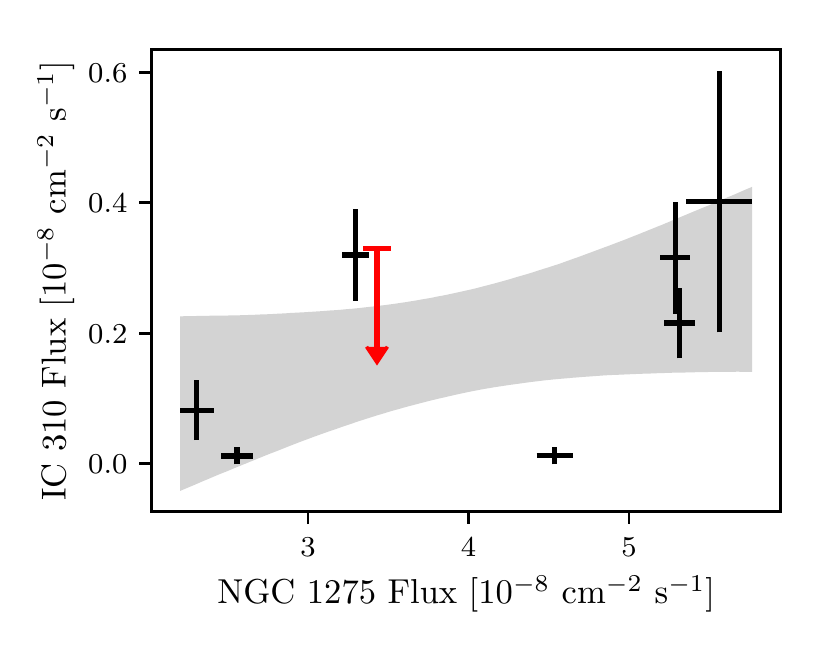}
\caption{Flux of NGC~1275 and IC~310 in each of the time periods studied in Section~\ref{sec:Temporal}. Fluxes are integrated between 100~MeV and 500~GeV, and error bars on points represent 1$\sigma$ uncertainty intervals. The grey area represents the 1$\sigma$ uncertainty period related to a linear fit to this data. Bins with a detection significance $<3\sigma$ are shown as upper limits.}
\label{fig:LC_NGC1275}
\end{figure*}

Many of the concerns in Section~\ref{sec:Systematics} relate to systematic bias on soft flux becoming associated with IC~310 due to the proximity of NGC~1275. If systematic errors in the \texttt{PSF} event classes used in this analysis were to wrongly associate low energy events (associated with a larger PSF) from NGC~1275 to IC~310 due to inaccurate PSF reconstruction, any increase in flux in NGC~1275 would result in an increase in flux in IC~310. Figure~\ref{fig:LC_NGC1275} shows the flux of both sources in each temporal bin used in Section~\ref{sec:Temporal}. These data show no significant correlation in the soft-spectrum bins of IC~310 in which we claim a detection. A linear fit to these data of the form $\text{Flux}_{\text{IC310}} = \alpha + \beta \cdot \text{Flux}_{\text{NGC1275}}$ with best-fit parameters $\alpha=(-0.03\pm0.33)\times10^{-8}$~ph~cm$^{-2}$~s$^{-1}$ and $\beta=(0.053\pm0.080)$, fully consistent with no dependence. 

Over the 8-year period considered in our analysis, NGC~1275 has increased in flux by a factor of two, and yet we find only an upper limit in our analysis of IC~310. This behaviour is inconsistent with a systematic error introduced by our use of the \texttt{PSF} instrument response functions mis-attributing flux between objects. A log-parabolic model of NGC~1275 was used for this investigation and spectrally fit in each bin. It is possible that deviations from this phenomenological model could provide a further systematic concern, but this has not been previously observed in NGC~1275. Indeed, many studies \citep[e.g.]{Manuel} assume that there is no significant spectral deviation from a log parabola in 8 years of integrated data.

%%%%%%%%%%%%%%%%%%%%%%%%%%%%%%%%%%%%%%%%%%%%%%%%%%

% Don't change these lines
\bsp	% typesetting comment
\label{lastpage}
\end{appendix}
\end{document}